%% file: main.tex
\documentclass[10pt,conference,final]{IEEEtran}

\usepackage[noadjust]{cite}

\usepackage{graphicx} 
\usepackage{physics}
\usepackage{amsmath}
\usepackage{amssymb}
\usepackage{balance}
\usepackage[hidelinks,colorlinks=true,linkcolor=blue,citecolor=blue]{hyperref}

\usepackage{tikz}
\usetikzlibrary{quantikz}
\usepackage{adjustbox}

\usepackage{caption}
\usepackage{subcaption}

\begin{document}
\title{\LARGE 
{\bf
Quantum Benchmarking via Random Dynamical Quantum Maps
}
\thanks{This work was partially supported by the NSF grant CCF-1908131.}
}

\author{
\IEEEauthorblockN{Daniel Volya and Prabhat Mishra}
\IEEEauthorblockA{University of Florida, Gainesville, Florida, USA}
}
\maketitle

\begin{abstract}
We present a benchmarking protocol for universal quantum computers, achieved through the simulation of random dynamical quantum maps.
This protocol provides a holistic assessment of system-wide error rates, encapsulating both gate inaccuracies and the errors associated with mid-circuit qubit measurements and resets.
By employing random quantum circuits and segmenting mid-circuit qubit measurement and reset in a repeated fashion, we steer the system of qubits to an ensemble of steady-states.
These steady-states are described by random Wishart matrices, and align with the steady-state characteristics previously identified in random Lindbladian dynamics, including the universality property.    
The protocol assesses the resulting ensemble probability distribution measured in the computational basis, effectively avoiding a tomographic reconstruction.
Our various numerical simulations demonstrate the relationship between the final distribution and different error sources.
Additionally, we implement the protocol on state-of-the-art transmon qubits provided by IBM Quantum, drawing comparisons between empirical results, theoretical expectations, and simulations derived from a fitted noise model of the device.
    
\end{abstract}


\section{Introduction}

\begin{figure}[tp]
    \centering
    \includegraphics[width=0.7\linewidth]{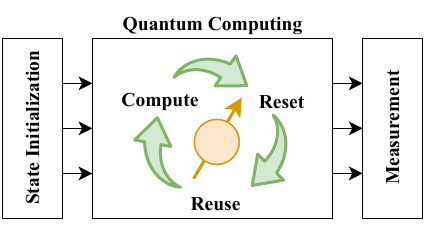}
    \vspace{-0.05in}
    \caption{\textbf{Quantum computing model.} Qubits are first initialized to a known fiducial state. Quantum information is then processed via a sequence of discrete gates acting on the qubits. These qubits may be reset, reinitialized, and reused throughout the computation, such as in an error correcting code or for post-selection. The final state is then measured. }
    \label{fig:quantum_computing}
\vspace{-0.2in}
\end{figure}

Quantum computers offer promising theoretical advantages over their classical counterparts, such as  prime factoring \cite{shorPolynomialTimeAlgorithmsPrime1997, monzRealizationScalableShor2016} and  search \cite{groverFastQuantumMechanical1996}. 
However, realizing quantum computers remains a difficult engineering challenge.
Most importantly, quantum computers must satisfy the following contradictory requirements with sufficient degrees of accuracy \cite{divincenzoPhysicalImplementationQuantum2000}: state initialization, single-and-multiple qubit gates, and individual qubit measurement.
While we desire granular control of individual qubits, we also seek to isolate the qubits from unwanted environmental degrees-of-freedom.
In recent years, we have witnessed considerable experimental progress, marked by the development and stabilization of many-qubit systems with decent accuracy \cite{aruteQuantumSupremacyUsing2019a}.
Given the scale and error rates, it is now feasible to algorithmically correct hardware errors via quantum error-correcting codes \cite{acharyaSuppressingQuantumErrors2023, bravyiHighthresholdLowoverheadFaulttolerant2024,dasilvaDemonstrationLogicalQubits2024}.
However, realizing error-correcting codes further requires real-time (mid-circuit) qubit measurement and qubit reset for syndrome extraction and correction, as shown in Fig.~\ref{fig:quantum_computing}.
In order to continue making steps towards fault-tolerant quantum computers, it is therefore crucial to develop benchmarks incorporating all the key ingredients.

Understanding the overall behavior of the system emerges as a significant and challenging question.
Although essential, characterization techniques such as state \cite{smolinEfficientMethodComputing2012} and process tomography \cite{poyatosCompleteCharacterizationQuantum1997, kneeQuantumProcessTomography2018a}, gateset tomography \cite{nielsenGateSetTomography2021a}, and measurement tomography \cite{chenDetectorTomographyIBM2019a}, are infeasible for systems larger than a handful of qubits.
State-of-the-art scalable methods such as randomized benchmarking \cite{dankertExactApproximateUnitary2009, gambettaCharacterizationAddressabilitySimultaneous2012} provide average error rates of gates by conducting long sequences of random gates, but do not capture system-wide properties.  
Metrics such as quantum volume \cite{crossValidatingQuantumComputers2019} are linked with system error rates and provide a quantification to the effective number of qubits of a quantum computer.
Unfortunately, these metrics do not capture properties such as mid-circuit qubit measurement and qubit reset which are necessary for achieving error-correcting codes.
Alternatively, one may observe the performance of many small quantum workloads via an application benchmark, such as quantum Fourier transform and phase estimation, and obtain an estimate to the performance \cite{lubinskiQuantumAlgorithmExploration2024}.
While these application benchmarks may be suitable in the future, they are biased to the choice of workloads and may be difficult to debug.

We seek to address this issue by introducing a protocol based on the steady-state properties of random open quantum system (OQS) dynamics.
In contrast to closed quantum systems -- where the effects of randomness have been thoroughly investigated, exemplified by the quantum chaos conjecture \cite{bohigasCharacterizationChaoticQuantum1984} and the characterization of system dynamics through a Poisson distribution \cite{aruteQuantumSupremacyUsing2019a,berryLevelClusteringRegular1997} -- the exploration of dynamics within random open quantum systems continues to be an area of active development.
Prior research has concentrated on unstructured discrete-time quantum maps \cite{bruzdaRandomQuantumOperations2009,bruzdaUniversalitySpectraInteracting2010, szehrDecouplingUnitaryApproximate2013} and the impact of decoherence from a stochastic environment \cite{gorinDecoherenceChaoticIntegrable2003,gorinRandomMatrixTheory2008}, leveraging foundational studies on non-Hermitian Hamiltonians \cite{sokolovStatisticalTheoryOverlapping1988,sokolovDynamicsStatisticsUnstable1989}.
Lately, attention has turned towards the continuous-time random Lindblad dynamics of Markovian open quantum systems \cite{canRandomLindbladDynamics2019, canSpectralGapsMidgap2019, saSpectralSteadystateProperties2020, wangHierarchyRelaxationTimescales2020}.
Remarkably, the steady-states of these random Lindbladian dynamics has been identified as being universal \cite{denisovUniversalSpectraRandom2019, hamazakiUniversalityClassesNonHermitian2020, akemannUniversalSignatureIntegrability2019}.
Recent works have also connected properties of continuous models to those discrete-time quantum maps \cite{saSpectralTransitionsUniversal2020,volya2024St}, including the property of universal steady-states.
We take advantage of the properties of discrete quantum maps to develop a quantum benchmark.

\begin{figure}[tp]
    \centering
    \begin{subfigure}[b]{0.72\linewidth}
        \centering
        \include{tikz_figs/circuit}
        \vspace{-10pt}
        \caption{}
    \end{subfigure}
    \hfill
    \begin{subfigure}[b]{0.25\linewidth}
        \centering
        \includegraphics[width=\linewidth]{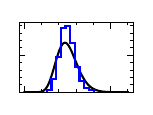}
        \caption{}
    \end{subfigure}
    \hfill
    \begin{subfigure}[b]{1\linewidth}
        \vspace{5pt} 
        \include{tikz_figs/channel}
        \vspace{-10pt}
        \caption{\label{fig:random-circuit-layers}}
    \end{subfigure}
    \caption{\textbf{Circuit model.} (a) A random dynamical map $\mathcal{E}$ is repeated $t$ times, mapping arbitrary initial states $\rho$ to a steady-state $\rho_t$. (b) Measurement of $\rho_t$ for an ensemble of random quantum maps produces a distinguishable distribution (blue). Deviations from the ideal distribution (black) correspond to system errors. (c) The map $\mathcal{E}$ is constructed via a random circuit of depth $d$, concluding with a measurement and reset.}
    \label{fig:random-circuit}
\vspace{-0.2in}
\end{figure}

In Fig.~\ref{fig:random-circuit}, we describe our approach involving circuits that are structured through the sequential layering of pairwise random gates, coupled with the measurement and reset of qubits.
Each layer comprises random gates and measurements, together facilitating an approximate-random, completely positive and trace-preserving map, symbolized as $\mathcal{E}$.
The repetitive application of this standardized layer, denoted as $\mathcal{E}^t$, gradually transitions any initial state $\rho$ towards a steady-state $\rho_{\mathrm{ss}}$ governed by the spectral characteristics of $\mathcal{E}$,
\begin{equation}
    \mathcal{E}^t(\rho) = \rho_t \approx \rho_{\mathrm{ss}}.  
\end{equation}
By generating an ensemble of random maps $\mathcal{E}$, we obtain an ensemble of steady-states, forming a measurable probability distribution that serves as a metric for evaluating the performance of the device. 

This paper makes the following major contributions.

\begin{itemize}
\item Establishes the properties of discrete quantum maps, with an emphasis on the random ensemble of these maps and examines their theoretical spectral properties.
\item Explores various strategies for the numerical generation of random quantum maps and discuss the construction of quantum circuits to actualize these maps.
\item Conducts a numerical investigation into how different sources of error impact the theoretical properties.
\item Experimental evaluation using two IBM Quantum computers (\texttt{ibm\_tokyo} and \texttt{ibm\_osaka}) and benchmarking their performance.
\end{itemize}

The remainder of this paper is organized as follows. Section~\ref{sec:approach} describes our quantum benchmarking scheme. Section~\ref{sec:results} presents experimental results on IBM quantum computers. Finally, Section~\ref{sec:conclusion} concludes the paper.

\section{Random Dynamical Quantum Maps}
\label{sec:approach}

In this paper, we focus on quantum dynamical maps that are completely positive and trace preserving (CPTP), also referred to as quantum channels.
The canonical representation of a quantum map $\mathcal{E}$ is via the operator-sum representation
\begin{equation}
    \mathcal{E}(\rho) = \sum^r_{i=1} K_i \rho K_i^\dagger
\end{equation}
where $r$ denotes the Kraus rank and $K_i$ are Kraus operators that satisfy the trace-presevation constraint $\sum_i K_i^\dagger K_i = \mathbb{I}$.
For $n$-qubits, the max Kraus rank is $r = 4^n$ which captures the full dimension of the Hilbert space \cite{choiCompletelyPositiveLinear1975a}.
The successive applications of the quantum map on an initial state $\rho$ results in a final state $\rho_t = \mathcal{E}^t(\rho) = \mathcal{E}\circ\mathcal{E}\hdots \circ\mathcal{E}(\rho)$.

\begin{figure*}[htp]
    \centering
    \begin{subfigure}[b]{0.25\linewidth}
         \centering
        \includegraphics[width=\linewidth]{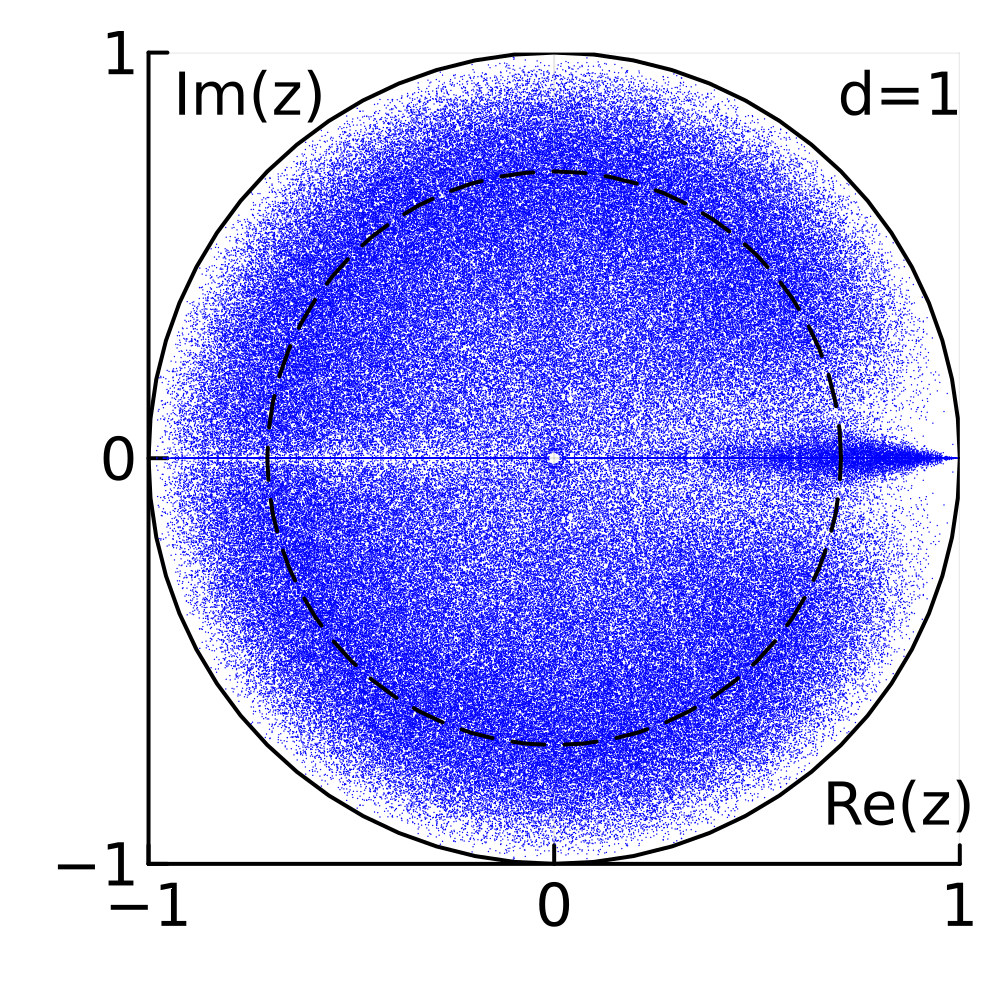}
        \vspace{-0.3in}
     \end{subfigure}
     \hfill
    \begin{subfigure}[b]{0.25\linewidth}
        \centering
        \includegraphics[width=\linewidth]{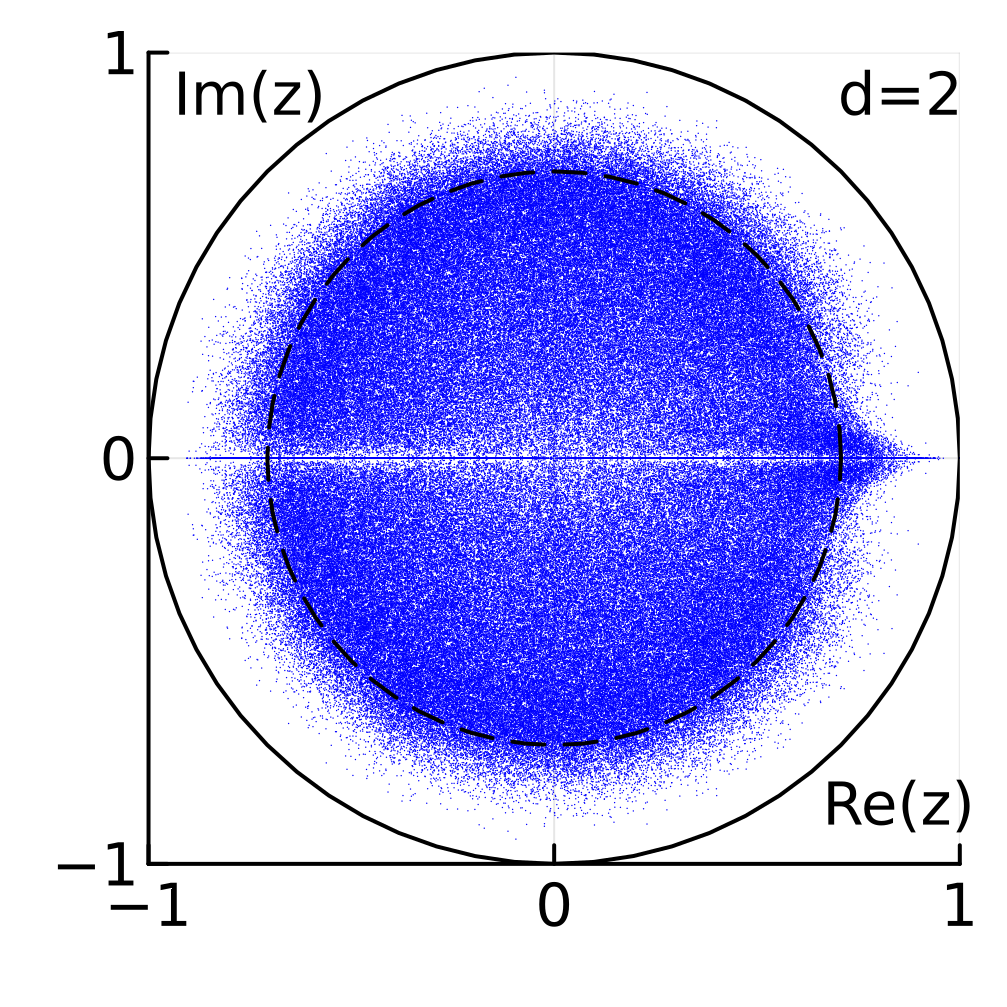}
        \vspace{-0.3in}
     \end{subfigure}
     \hfill
    \begin{subfigure}[b]{0.25\linewidth}
         \centering
        \includegraphics[width=\linewidth]{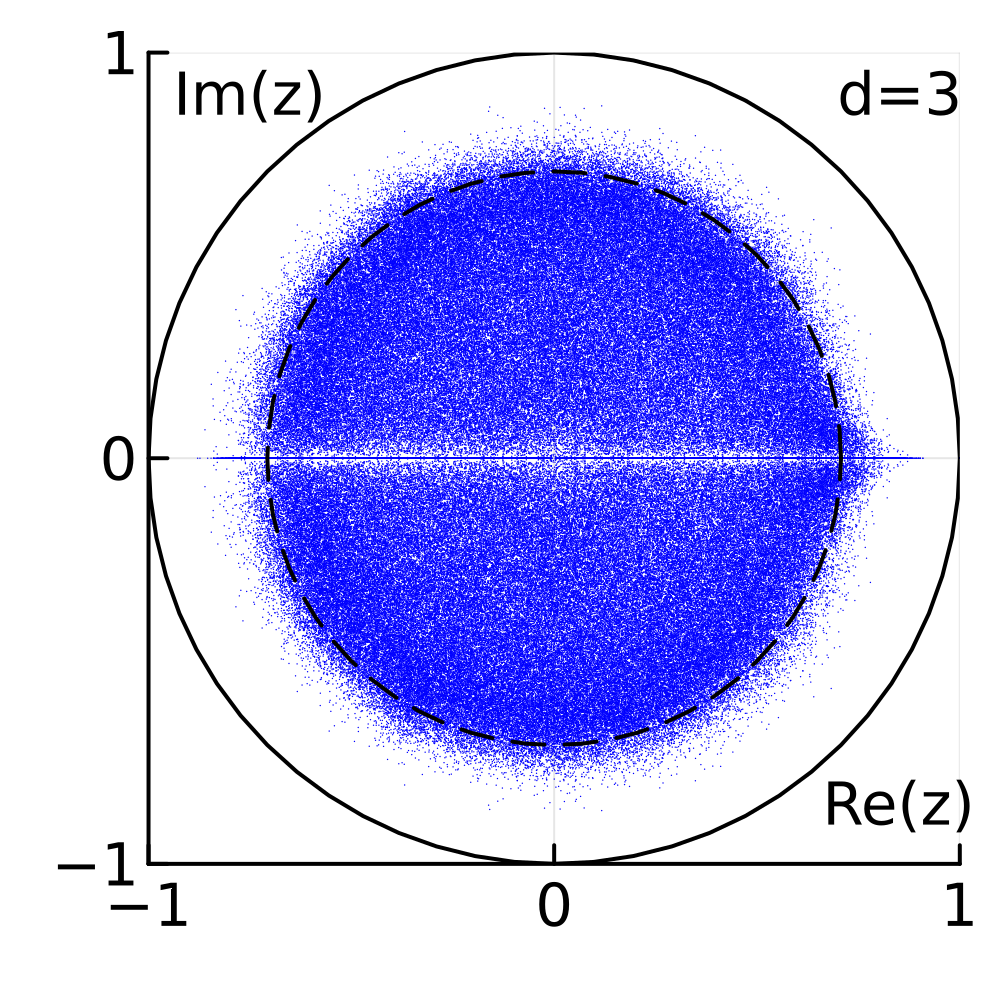}
        \vspace{-0.3in}
     \end{subfigure}
     \vspace{-0.03in}
     \caption{Eigenvalue spectrum of random dynamical maps, of rank $r=2$, formed by layering random two-qubit gates. The spectrum includes a leading eigenvalue $\lambda=1$ and the theoretical Girko disk of radius $1/\sqrt{r}$ for the remaining eigenvalues $\lambda$. With successive layering $d$ of random gates, the empirical eigenvalues $\lambda$ better approximate the Girko disk.  }
    \label{fig:map_eigenvalues}
\end{figure*}

This section is organized as follows. We first consider the spectral properties of quantum maps. Next, we discuss the natural candidates for probability distribution on the set of quantum maps, including their numerical generation.
Then, we examine the eigenvalue spectrum of an ensemble of random maps, which determines the convergence rate to a steady-state.
Next, we investigate the ensemble steady-state spectrum to which the system converges to by the application of random quantum maps.
We then examine the statistics obtained by measuring the final ensemble of steady-states.
Finally, we construct random quantum circuits to approximate a random dynamical map.

\subsection{Spectral Properties of Quantum Maps}
Given a sufficient number of iterations $t$, the final state will converge to a steady-state.
The steady-state can be analyzed by the spectral decomposition of $\mathcal{E}$, namely the eigenpair relation
\begin{equation}
    \mathcal{E}(\rho_\lambda) = \lambda \rho_\lambda
\end{equation}
where $\lambda$ is the eigenvalue corresponding to the eigenvector $\rho_\lambda$.
The eigenvalues $\lambda$ are complex numbers, and due to the CPTP condition, they reside inside a unit disk.
In the spectrum of eigenvalues, there is at least one unit eigenvalue \cite{evansSpectralPropertiesPositive1978a}, $\lambda=1$, which we denote as a fixed point.
There may be eigenvalues that live on the edge of the unit disk, $|\lambda|=1$, which are known as rotating points.
The associated eigenvectors of the fixed points and rotating points define steady-state space.
In this paper, the properties of our maps are such that there are no rotating points, hence the eigenvector associated with fixed point is equal to the converging steady-state $\rho_{\mathrm{ss}}$.
By constructing an ensemble of random maps, we obtain an ensemble of random steady-states.

\subsection{Probability Distribution of Random Quantum Maps}\label{sec:rand_maps}

To numerically study random quantum maps, we pick Kraus operators $K_i$ at random by constructing Ginibre matrices $G_i$ where each element consists of independent Gaussian entries.
We define $K_i = G_i S^{-1/2}$ where $S$ is a normalization factor given as $S = \sum_i G_i^\dagger G_i$.
We note that this method is the fastest procedure for generating a random quantum map.
Alternatively, one may consider the following natural options for probability distributions on the convex set of quantum dynamical maps:
(1) Lebesgue measure, (2) Haar-random isometry $W$ in the Stinespring decomposition, and (3) random Choi matrix.
Under correct parameters, these methods are all equivalent \cite{kukulskiGeneratingRandomQuantum2021}.
For theoretical investigations in this paper, it is insightful to consider the Stinespring decomposition (2), as it has direct correspondence with the random circuits in Fig.~\ref{fig:random-circuit-layers} and in Sec.~\ref{sec:circuits}.
In particular, we have the following relationship

{
\vspace{-0.1in}
\begin{align}\label{eq:stinespring_kraus}
    \mathcal{E}(\rho) &= \mathrm{Tr}_\mathrm{env}\left[U(\rho \otimes \rho_\mathrm{env})U^\dagger\right] \nonumber \\
    &= \sum_{i=1}^r \bra{e_i} U [\rho \otimes \ket{e_0}\bra{e_0}]U^\dagger \ket{e_i} = \sum_{i=1}^r K_i \rho K_i^\dagger
\end{align}
where tracing over $r$ environmental degrees-of-freedom is equivalent to the Kraus representation  of rank $r$. 
}

\subsection{Eigenvalue Spectrum of Random Quantum Maps}\label{sec:channel_spectrum}
As noted in \cite{kukulskiGeneratingRandomQuantum2021,bruzdaRandomQuantumOperations2009}, the properties of $\mathcal{E}$ corresponding to a random map can be modeled by the real Ginibre ensemble.
In particular, the second-largest eigenvalue $|\lambda_2| \leq 1$  lives inside the Girko disk of radius $\gamma = 1/\sqrt{r}$ where $r$ is the rank of the map.
The spectral gap, $\Delta = 1 - \gamma \geq 0$, then determines the convergences of the system to a steady-state $\rho_\mathrm{ss}$.
Namely, a state $\rho_t$ is close to the steady-state after a $l/\Delta$ iterations, given an integer $l$.
Ideally, to achieve a desired tolerance $\epsilon$, the number of iterations $t$ required is
    $t \approx \frac{\log(\epsilon)}{\log(\gamma)}$.
Figure~\ref{fig:map_eigenvalues} visualizes the eigenvalues for an ensemble of random maps with rank $r=2$, and displaying the expected Girko disk of radius $1/\sqrt{2}$.

\begin{figure}[htp]
    \centering
    \includegraphics[width=0.85\linewidth]{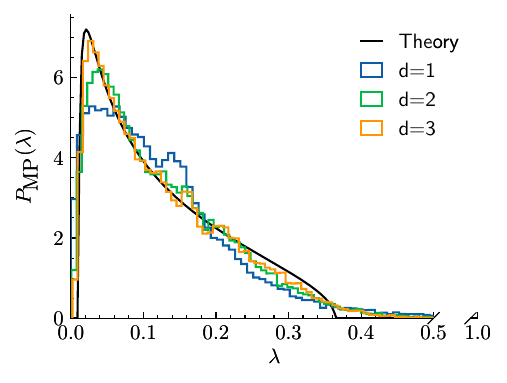}
    \vspace{-0.15in}
    \caption{Steady-state eigenvalue spectrum of random maps $\mathcal{E}$, corresponding to a rescaled Marchecko-Pastur probability distribution (theory). Eigenvalues are computed for steady-states of maps with $n=3$ qubits and with varying random circuit depth $d$.}
    \label{fig:Marchenko_pastur}
\end{figure}

\begin{figure*}[thp]
    \centering
    \begin{subfigure}[b]{0.32\linewidth}
        \centering
        \includegraphics[width=1\linewidth]{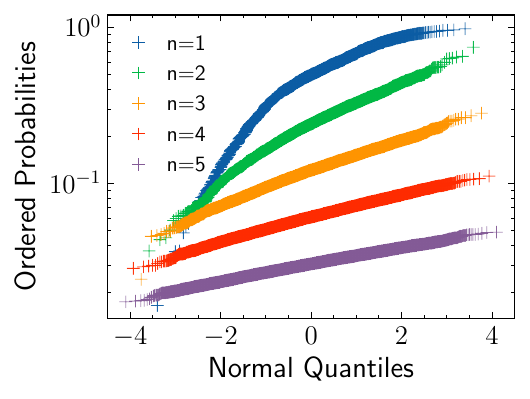}
        \vspace{-0.25in}
        \caption{}
    \end{subfigure}
    \hfill
    \begin{subfigure}[b]{0.32\linewidth}
        \centering
        \includegraphics[width=1\linewidth]{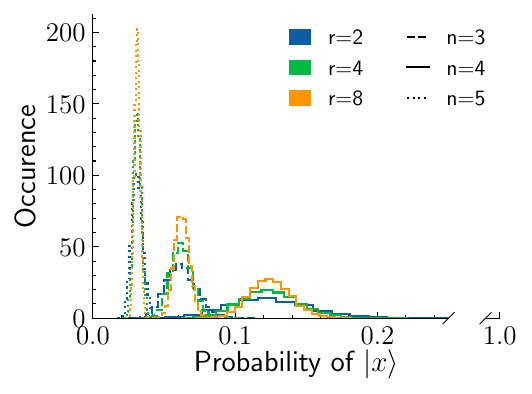}
        \vspace{-0.25in}
        \caption{}
    \end{subfigure}
    \hfill
    \begin{subfigure}[b]{0.32\linewidth}
        \centering
        \includegraphics[width=1\linewidth]{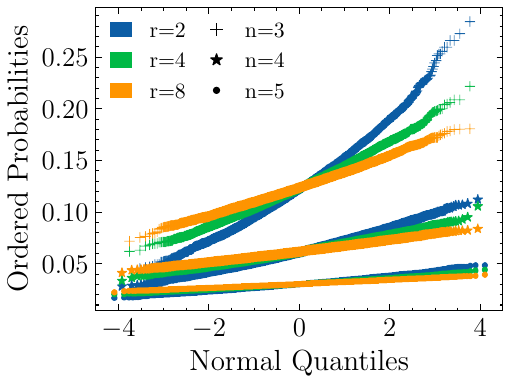}
        \vspace{-0.25in}
        \caption{}
    \end{subfigure}
    \vspace{-0.1in}
    \caption{Examination of output distribution of random circuits consisting of $n$-qubits. (a) Normal probability plot comparing output distributions to the theoretical normal distribution for  $n=\{1,2,3,4,5\}$-qubits (note the log-scale).  (b) Histogram of output distribution for $n=\{3,4,5\}$ of rank $r=\{2,4,8\}$. (c) Normal probability plot for histogram in (b).\label{fig:output-measurements}}
\end{figure*}

\subsection{Steady-state Spectrum}\label{sec:steady-state}

As discussed in Sec.~\ref{sec:channel_spectrum}, the repeated application of a random map $\mathcal{E}$ on an arbitrary state converges to a steady-state $\rho_\mathrm{ss}$.
The properties of the steady-state are understood through the eigen decomposition
\begin{equation}
    \rho_{\mathrm{ss}} = UDU^\dagger = \sum_i \lambda_{\mathrm{MP}} \ket{\psi_i}\bra{\psi_i}.
\end{equation}
Since any density matrix is a positive semidefinite Hermitian operator, it can be diagonalized through a unitary matrix $U$ and a diagonal matrix $D$ consisting of positive and real eigenvalues.
This decomposition is given as a mixture of random, pure quantum states, $\ket{\psi_i}$, weighted by the eigenvalues.
In this case, the eigenvalues $\lambda_{\mathrm{MP}}$ follow a Marchenko-Pastur distribution \cite{marcenkoDISTRIBUTIONEIGENVALUESSETS1967,edelmanEigenvaluesConditionNumbers1988,nadalStatisticalDistributionQuantum2011} as shown in Fig.~\ref{fig:Marchenko_pastur}.
This is consistent with the results observed in the entanglement spectrum of random bipartite systems \cite{lubkinEntropySystemIts1978,zyczkowskiInducedMeasuresSpace2001,sommersStatisticalPropertiesRandom2004,znidaricEntanglementRandomVectors2006,zyczkowskiGeneratingRandomDensity2011}. In these systems, a partial trace is executed over all environmental degrees-of-freedom after the combined system-environment undergoes an extended period of unitary dynamics evolution.
In particular, due to CPTP constraints, we must consider steady-states with $\mathrm{Tr}(\rho_\mathrm{ss}) = 1$.
By considering fixed-trace Wishart ensembles, we obtain a rescaled Marchenko-Pastur probability distribution \cite{nadalStatisticalDistributionQuantum2011}, 
\begin{equation}\label{eq:march}
    P_{\mathrm{MP}}(\lambda) = \frac{1}{2\pi\kappa}\frac{1}{\lambda}\sqrt{(\lambda_{+} - \lambda)(\lambda-\lambda_{-})}
\end{equation}
where $\kappa=1/(Nr)$ for Hilbert space $N$ ($N=2^n$) and
\begin{equation}
    \lambda_\pm = \frac{1}{N} \left(1\pm \frac{1}{\sqrt{r}} \right)^2. 
\end{equation}
The average of the $P_{\mathrm{MP}}(\lambda)$ is $1/N$, corresponding to a completely mixed state.
However, the variance $\sigma^2_\mathrm{MP}$ follows as $1/(N^2 r)$, with higher moments found as \cite{mingoFreeProbabilityRandom2017}
\begin{equation}
    \mu_m = \frac{1}{m}\frac{1}{Nr}\sum_{l=1}^{m} \binom{m}{l-1}\binom{m}{l} r^l.
\end{equation}

The eigenvectors $U$ ($\ket{\psi_i}$) are themselves Haar random, aligning to a random unitary ensemble.
Since the steady-state is a weighted mixture of an ensemble of random pure states, the steady-state is said to be universal, coinciding with prior results in the continuous Lindbladian regime \cite{denisovUniversalSpectraRandom2019}.

\subsection{Measurement Statistics of Steady-State}

Measurements of $\rho_\mathrm{ss}$ are given by the diagonal matrix elements.
For instance, the probability of $\ket{011}$ is given as $\bra{011}\rho_{\mathrm{ss}}\ket{011}$, corresponding to the magnitude-squared of the third diagonal matrix element, assuming the standard computational basis.
We denote the measurement bitvector as $\ket{x}$, hence the probability outcome of $\ket{x}$
\begin{equation}\label{eq:moments}
    \mathrm{Pr}_{\ket{x}} = \bra{x}\rho_{\mathrm{ss}}\ket{x} = \sum_i \lambda_{\mathrm{MP}_i} |\bra{\psi_i}\ket{x}|^2 = \sum_i \lambda_{\mathrm{MP}} \mathrm{PT}(n)
\end{equation}
where the magnitude-squared elements in an ensemble of random state vectors yields the Porter-Thomas distribution \cite{porterFluctuationsNuclearReaction1956, volyaPorterThomasDistributionUnstable2011}.
The probabilities of $\ket{x}$ are a result of a combined Marchenko-Pastur and Porter-Thomas (PT) distribution, and due to the invariance of the Haar measure, are equivalent irrespective of basis or choice of $\ket{x}$.
In other words, the diagonal elements of the steady-state $\rho_\mathrm{ss}$ are equivalent probability distributions.
We note that this invariance is expected as there is no favoritism in the configuration space, which satisfies the eigenstate thermalization hypothesis \cite{srednickiChaosQuantumThermalization1994}. Figure~\ref{fig:output-measurements} compares the results of $\mathrm{Pr}_{\ket{x}}$ for different qubit counts $n$, where $N=2^n$, and for different ranks $r$. These results illustrate the relationship described by (\ref{eq:moments}). As $N$ and $r$ increase, the distribution becomes narrower. For a fixed $N$, higher values of $r$ reduce the tails of the distribution. Comparisons to normal distribution quantiles visualize the loss of tails for increased $N$ and $r$. 

\subsection{Quantum Circuits for Quantum Maps}\label{sec:circuits}

By the Stinespring decomposition discussed in Sec.~\ref{sec:rand_maps}, we can realize random dynamical quantum maps by working in a larger Hilbert space and tracing away the ancilla (environment) degrees-of-freedom.
It then follows that a random dynamical map may be constructed by a composition of these key steps:
\begin{enumerate}
    \item initialize an ancilla to a known state, denoted as $\ket{0}$ without loss of generality,
    \item sample a random unitary $U$ for the combined system and ancilla,
    \item trace away the ancilla, which is equivalent to measuring the ancilla and taking an unbiased average.
\end{enumerate}
Following (\ref{eq:stinespring_kraus}), this procedure produces $r$ random Kraus operators $K_i$ which express our channel $\mathcal{E}$. 

\begin{figure*}[t]
    \centering
     \begin{subfigure}[b]{0.31\linewidth}
        \centering
        \includegraphics[width=1\linewidth]{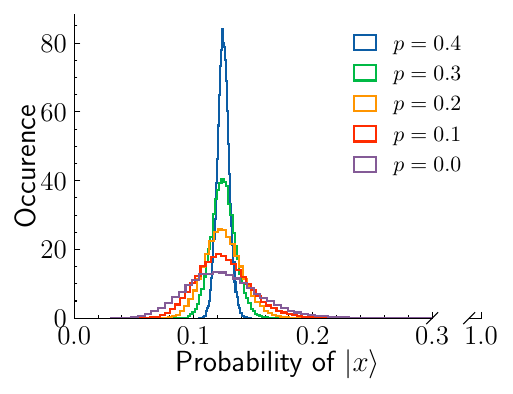}
        \vspace{-0.25in}
        \caption{}
    \end{subfigure}
    \hfill
    \begin{subfigure}[b]{0.31\linewidth}
        \centering
        \includegraphics[width=1\linewidth]{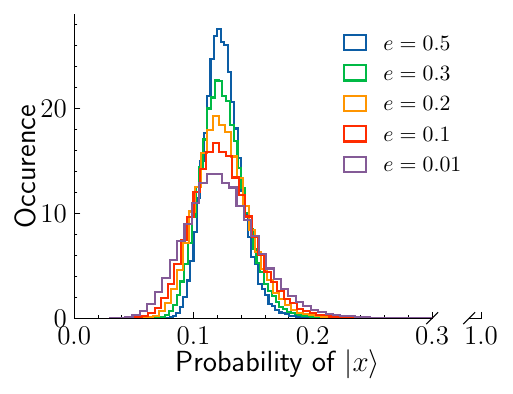}
        \vspace{-0.25in}
        \caption{}
    \end{subfigure}
    \hfill
    \begin{subfigure}[b]{0.31\linewidth}
        \centering
        \includegraphics[width=1\linewidth]{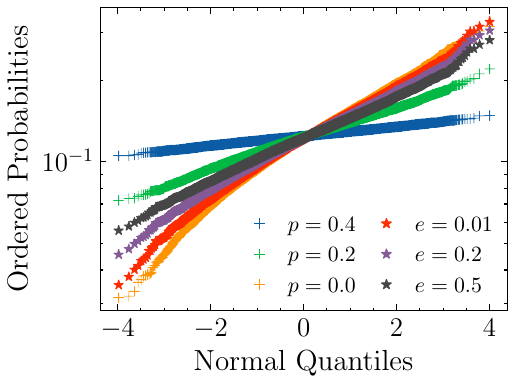}
        \vspace{-0.25in}
        \caption{}
    \end{subfigure}
    \vspace{-0.05in}
    \caption{Output distributions consisting of $n=3$ qubits and $d=3$ random circuit depth with different noise sources. (a) Faulty reset of ancilla qubit parameterized by $p$, with ideal reset to $\ket{0}$ corresponding to $p=0$ and with $p=1/2$ representing an equal mixture of $\ket{0}$ and $\ket{1}$. (b) Decoherence modeled by a depolarizing channel with strength $e$ applied to each gate. (c) A probability plot of the distributions with respect to normal distribution quantiles. \label{fig:noise}}
\vspace{-0.1in}
\end{figure*}

\begin{figure}[t]
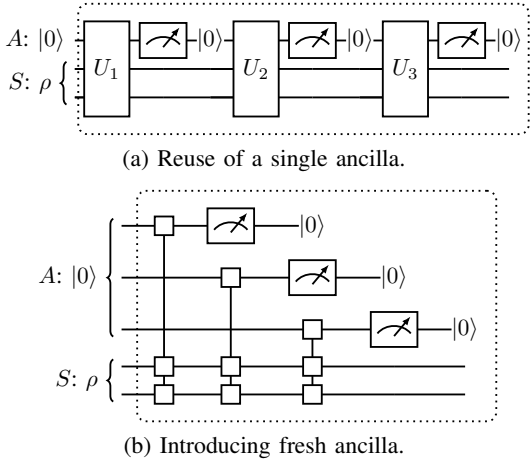

    \centering
    \vspace{-0.1in}
    \begin{subfigure}[b]{0.8\linewidth}
        \include{tikz_figs/circ_1}
        \vspace{-0.1in}
        \caption{Reuse of a single ancilla.\label{fig:ancilla_reuse}}
    \end{subfigure}
    \begin{subfigure}[b]{0.7\linewidth}
        \include{tikz_figs/circ_2}
        \vspace{-0.1in}
        \caption{Introducing fresh ancilla.\label{fig:fresh_ancilla}}
    \end{subfigure}
    \caption{External degrees-of-freedom may be introduced by either (a) reset and reuse of qubits, or (b) allocating fresh qubits. The intention of the protocol is to reuse the qubits.}
    \label{fig:ancilla}
    \vspace{-0.1in}
\end{figure}

In the circuit model of quantum computing, it is not possible to apply a global unitary $U$ acting simultaneously on all qubits.
Instead, a naive strategy is to apply gates from a universal set randomly, which will require an exponential number of gates in terms of $n$-qubits \cite{knillApproximationQuantumCircuits1995}.
In other words, sampling from the uniform Haar distribution is inefficient.
In our case, and in many cases, applying pseudo-random operators is sufficient.
The extent to which the pseudo-random operators behave like the uniform distribution is known as $k$-design (sometimes referred to as $t$-design).
A $k$-design has the $k$-th moment equal to those of Haar distribution \cite{ambainisQuantumTdesignsTwise2007, harrowRandomQuantumCircuits2009}.
However, methods for constructing unitary $k$-designs (such as using Clifford group) remain inefficient.
Instead, we may opt for random quantum circuits, where a circuit acting on $n$ qubits with length $poly(n,k)$ approximates a $k$-design.
For instance, it has been shown that a random circuit of length $O(n(n+\log 1/\epsilon))$ is an $\epsilon$ approximation for a $2$-design \cite{harrowRandomQuantumCircuits2009}.

Figure~\ref{fig:random-circuit-layers} is an example random circuit of depth $d$ acting on $4$-qubits.
With each layer $d$, the moments of the distribution are corrected.
After $d$ layers of pair-wise random gates, the ancilla is measured with the results averaged (ignored).
These steps produce a random dynamical map $\mathcal{E}$. Figure~\ref{fig:map_eigenvalues} shows the eigenvalues a channel $\mathcal{E}$ generated by a $d$-layered circuit. As the depth $d$ is increased, we see agreement in the non-dominant eigenvalues residing in the Girko disk. After the ancilla is measured, it is reset to a known state so that we can repeat the same random quantum circuit ($\mathcal{E}$).
Upon several repetitions of $\mathcal{E}$, following the analysis of Sec.~\ref{sec:steady-state}, the system converges to a steady-state governed by a Marchenko-Pastur distribution. Fig.~\ref{fig:Marchenko_pastur} highlights the distribution for different circuit depths $d$.

While the intention of the protocol is to benchmark mid-circuit reset, one may instead opt to use freshly prepared ancillas after each iteration as depicted in Fig.~\ref{fig:ancilla}.
However, this may be costly to perform, requiring additional swap operations to move fresh qubits to couple with the system qubits.

\subsection{Effect of Noise}

We numerically analyze two simple sources of noise:
\begin{enumerate}
    \item {a faulty reset of the ancilla qubit, modeled as
    \begin{equation}
        \rho(p) = \sqrt{p}\ket{0}\bra{0} + \sqrt{1-p}\ket{1}\bra{1}; \; p \in [0,0.5]
    \end{equation}
    }
    \item {depolarizing error of the state after a gate, modeled as
    \begin{equation}
        \rho \mapsto \frac{(1-e)}{N}\mathbb{I} + e\rho; \; e\in[0,0.5]
    \end{equation}
    }
\end{enumerate}
As shown in Fig.~\ref{fig:noise}, we see that both sources of noise lower the variance of the measurement distribution.
In particular, both the error in the reset of the ancilla qubit and the error via a depolarizing channel result in a loss of quantum information, converging to a completely mixed quantum state.
The rank of the channel increases, resulting in a new Marchenko-Pastur distribution (\ref{eq:march}). 
Coherent and unbiased error in quantum gates will, on average, have no impact on the output statistics. This is a result of the invariance of the Haar measure, where a change in basis does not affect the statistics.
On the other hand, biased sources of noise will place preference to certain states, leading to a change in the distribution.
This includes for example a decay channel, or unequal Pauli-noise channels.

\begin{figure*}[t]
    \centering
    \begin{subfigure}[b]{0.45\linewidth}
        \centering
    \includegraphics[width=1\linewidth]{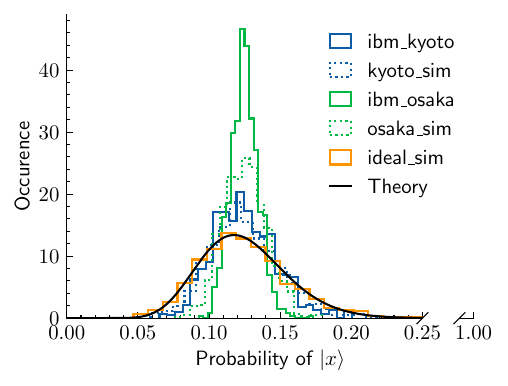}

    \end{subfigure}
    \hfill
    \begin{subfigure}[b]{0.45\linewidth}
        \centering
        \includegraphics[width=1\linewidth]{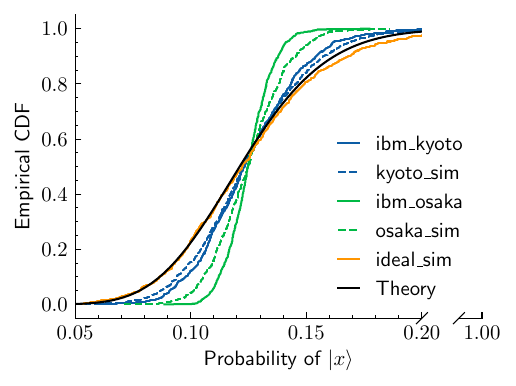}
    \end{subfigure}
    \vspace{-0.1in}
    \caption{Benchmark results on IBM Quantum computers \texttt{ibm\_kyoto} and \texttt{ibm\_osaka}, along with noisy simulations, and an ideal simulation. A circuit consists of $D=3$ blocks with pair-wise two-qubit random gates, with a total of $N=10$ repetitions. The output probabilities were collected for an ensemble of $100$ random circuits. \label{fig:results}}
    \vspace{-0.1in}
\end{figure*}

\begin{figure}
    \centering
    \includegraphics[width=0.93\linewidth]{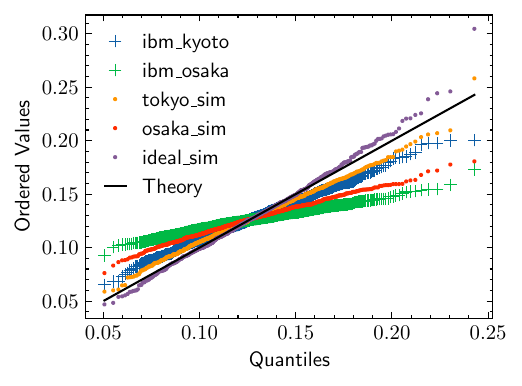}
    \vspace{-0.1in}
    \caption{Probability plot comparing device and noisy simulation results in terms of theoretical quantiles. Ideally, points should align with a line of slope $=1$. \label{fig:qq}}
    \vspace{-0.2in}
\end{figure}

\begin{table}[]
    \setlength{\tabcolsep}{3pt}
    \centering
    \caption{Qubit properties \cite{IBMQuantum} and the final Kolmogorov-Smirnov (KS) distance between empirical and expected result. Asterisk (*) denotes the ancilla qubit. The final column $p(1|0)$ is the error rate of resetting a qubit to $\ket{0}$.\label{tab:callibration}}
    \begin{tabular}{lr|crrrr}
    \hline\hline
    \textbf{Device}     & \textbf{KS}     & \textbf{Qubit} & \textbf{T1 [\textmu s]} & \textbf{T2 [\textmu s]} & \textbf{Readout} & $\mathbf{p(1|0)}$ \\ \hline\hline
    \texttt{ibm\_kyoto} & \textbf{0.103}  & 43    & 408.8 & 388.2 & 0.014   & 0.013  \\
                        &                 & 44    & 299.1 & 225.9 & 0.008   & 0.008  \\
                        &                 & 45    & 273.9 & 413.2 & 0.008   & 0.006  \\ 
                        &                 & *46   & 323.4 & 457.7 & 0.009   & 0.008  \\\hline
    \texttt{ibm\_osaka} & \textbf{0.275}  & *25   & 212.5 & 382.5 & 0.004   & 0.002  \\
                        &                 & 26    & 297.5 & 15.8  & 0.008   & 0.005  \\
                        &                 & 27    & 399.7 & 159.6 & 0.014   & 0.015  \\
                        &                 & 28    & 256.5 & 31.1  & 0.007   & 0.009  \\ \hline
    \end{tabular}
\vspace{-0.1in}
\end{table}

\section{Demonstrations on Quantum Computers}
\label{sec:results}

We demonstrate the protocol on two superconducting quantum computers from IBM Q, \texttt{ibm\_kyoto} and \texttt{ibm\_osaka} \cite{IBMQuantum}.
Additionally, we simulate the circuits using fitted noise models from the respective devices via Qiskit ecosystem \cite{Qiskit}.
We perform the protocol with $4$ qubits in total, with $1$ qubit acting as an ancilla qubit that is measured and reset. We conduct experiments on 100 randomly generated dynamical maps, constructed with random circuits of depth $d=3$, and executed with $4096$ shots. Raw data and code is available \cite{volyaRustyBambooRandomMaps2024}.

The error models of the devices include: (a) single and two qubit errors consisting of a depolarizing channel, (b) a thermal relaxation channel specified via  $T_1$ and $T_2$ relaxation time, and (c) a readout error.
The parameters of the error models are provided by the IBM Q team \cite{IBMQuantum}, and are summarized in Tab.~\ref{tab:callibration}.
A notable difference between the two devices is in the $T_2$ time, with \texttt{ibm\_kyoto} having an order of magnitude longer coherence time.

Figure~\ref{fig:results} illustrates both the resulting distribution and the cumulative distribution function (CDF) for the final output of the devices and simulations. We calculate the Kolmogorov-Smirnov (KS) distance -- the maximum vertical discrepancy between the empirical and theoretical CDFs -- to quantify the overall error. The KS distances for \texttt{ibm\_kyoto} paired with the simulation are 0.103 and 0.081, respectively, while \texttt{ibm\_osaka} paired with the simulation display distances of 0.275 and 0.149. Figure~\ref{fig:qq} further compares the distributions using a probability plot, highlighting the differences.
The observed discrepancies between the simulations and the actual device outputs are attributable to the simplifications made in the error models. These models primarily assume that all gate and measurement errors are local and Markovian, thereby excluding non-Markovian influences such as cross-talk, leakage, or drift. Furthermore, the models consider gate errors to arise solely from incoherent noise processes, which may not capture all sources of error accurately.

\section{Conclusions and Future Outlook}
\label{sec:conclusion}

In this work, we introduced a benchmarking protocol for quantum computers utilizing random dynamical quantum maps to simulate an open quantum system undergoing non-unitary dynamics. The protocol captures intrinsic errors from qubit gates, measurements, and resets through sequences of random gates, steering the system towards a steady-state characterized by random Wishart matrices. Our theoretical analysis, numerical simulations, and empirical validation on IBM Quantum computers demonstrated that the protocol's measure of quantum system performance based on the steady-state distribution of open quantum systems dynamics aligns with theoretical predictions and practical expectations.

There are numerous promising avenues for future research.
Additional research is required to establish the relationship between alternative benchmarks and the performance of error-correcting codes.
Another important avenue is further analysis of various noise sources, including the role of non-Markovian processes.
Furthermore, it is interesting to understand the role of random dynamical quantum circuits, where gates change depending on the outcome of mid-circuit measurement.
We believe that such future avenues, and our present work, provide a promising path to characterize quantum systems and their suitability for error correction and fault tolerance.

\section*{Acknowledgments}
We acknowledge the use of IBM Quantum services for this work. The views expressed are those of the authors, and do not reflect the official policy or position of IBM.
The computing systems used in this work is supported by DOE (DE-SC-0009883).
This research was partially supported by the grants from DARPA (HR0011-24-3-0004) and NSF (CCF-1908131).

\balance
\bibliographystyle{IEEEtran}
\bibliography{references}

\end{document}

%% file: tikz_figs/circuit.tex
\tikzset{
    slice/.append style={blue},
}
\begin{adjustbox}{width=1\linewidth}
\begin{quantikz}[column sep=8pt, row sep=2pt]
\lstick[3]{$\rho$} & \gate[wires=3]{\mathcal{E}(\rho)}  & \qw & \gate[wires=3]{\mathcal{E}(\mathcal{E}(\rho))} & \qw & \ldots\ & & \gate[wires=3]{\mathcal{E}^t(\rho)}\slice{$\rho_t$} & \meter[style={scale=1}]{}\\
& \qw  & \qw & \qw & \qw & \ldots\ & & \qw & \meter[style={scale=1}]{}\\
& \qw  & \qw & \qw & \qw & \ldots\ & & \qw & \meter[style={scale=1}]{}
\end{quantikz}
\end{adjustbox}

%% file: tikz_figs/channel.tex
\tikzset{
    slice/.append style={blue},
}
\begin{adjustbox}{width=1\linewidth}

\begin{quantikz}[column sep=4pt, row sep={20pt,between origins}]
\lstick{$\ket{0}$} & \gate[wires=4]{\mathcal{E}(\rho)} & \qw & \\
\lstick[3]{$\rho$}& \qw & \qw & \\
& \qw & \qw &\\
& \qw & \qw &
\end{quantikz}
=
\begin{quantikz}[column sep=3pt, row sep = 2pt]
    \lstick{$\ket{0}$} & \gate[2, style={fill=white}, label style={black}]{\mathrm{SU(4)}}\gategroup[wires=4, steps=2, style={rounded corners, fill=blue!20, inner xsep=0pt, inner ysep=0pt}, label style={label position=below, yshift=-0.5cm}, background]{$1$} & \qw & \qw & \ldots\ & \gate[2, style={fill=white}, label style={black}]{\mathrm{SU(4)}}\gategroup[wires=4, steps=2, style={rounded corners, fill=blue!20, inner xsep=0pt, inner ysep=0pt}, label style={label position=below, yshift=-0.5cm}, background]{$d$} & \qw & \qw & \meter{} & \push{\ket{0}} & \qw\\
    \lstick[3]{$\rho$} & & \gate[2, style={fill=white}, label style={black}]{\mathrm{SU(4)}} & \qw & \ldots\ & \qw & \gate[2, style={fill=white}, label style={black}]{\mathrm{SU(4)}} & \qw & \qw\\
    & \gate[2, style={fill=white}, label style={black}]{\mathrm{SU(4)}} \qw & \qw & \qw &  \ldots\ & \gate[2, style={fill=white}, label style={black}]{\mathrm{SU(4)}} & \qw & \qw & \qw\\
    & & \qw & \qw & \ldots\ &\qw & \qw & \qw & \qw
\end{quantikz}
\end{adjustbox}

%% file: tikz_figs/circ_1.tex
\tikzset{
    slice/.append style={blue},
}
\begin{adjustbox}{width=1\linewidth}

\begin{quantikz}[row sep=4pt, column sep=4pt]
    \lstick{$A$: $\ket{0}$} & \gate[3, style={fill=white}, label style={black}]{U_1}\gategroup[wires=3, steps=9, style={dotted, rounded corners, inner xsep=-1pt}, background]{} & \meter{}  & \push{\ket{0}} & \gate[3, style={fill=white}, label style={black}]{U_2} & \meter{}  & \push{\ket{0}} & \gate[3, style={fill=white}, label style={black}]{U_3} & \meter{}  & \push{\ket{0}} \\
    \lstick[2]{$S$: $\rho$} & \qw & \qw & \qw & \qw & \qw & \qw & \qw & \qw & \qw \\
    & \qw & \qw & \qw& \qw & \qw & \qw& \qw & \qw & \qw
\end{quantikz}
\end{adjustbox}

%% file: tikz_figs/circ_2.tex
\tikzset{
    slice/.append style={blue},
}
\begin{adjustbox}{width=1\linewidth}

\begin{quantikz}[row sep=4pt]
    \lstick[3]{$A$: $\ket{0}$} & \gate{}\vqw{3}\gategroup[wires=5, steps=5, style={dotted, rounded corners}, background]{} & \meter{} & \push{\ket{0}} \\
    & \qw & \gate{}\vqw{2} & \meter{} & \push{\ket{0}} \\
    & \qw & \qw & \gate{}\vqw{1} & \meter{} & \push{\ket{0}} \\
    \lstick[2]{$S$: $\rho$} & \gate{}\vqw{1} & \gate{}\vqw{1} & \gate{}\vqw{1} & \qw & \qw \\
    & \gate{} & \gate{} & \gate{} & \qw & \qw 
\end{quantikz} 
\end{adjustbox}